# RAPID: Contention Resolution-based Random Access using Context ID for IoT

Junseok Kim, *Student Member, IEEE*, Seongwon Kim, *Member, IEEE*,
Tarik Taleb, *Senior Member, IEEE*, and Sunghyun Choi, *Fellow, IEEE*

*Abstract*—With the increasing number of Internet of Things (IoT) devices, Machine Type Communication (MTC) has become an important use case of the Fifth Generation (5G) communication systems. Since MTC devices are mostly disconnected from Base Station (BS) for power saving, random access procedure is required for devices to transmit data. If many devices try random access simultaneously, preamble collision problem occurs, thus causing latency increase. In an environment where delay-sensitive and delay-tolerant devices coexist, the contention-based random access procedure cannot satisfy latency requirements of delay-sensitive devices. Therefore, we propose RAPID, a novel random access procedure, which is completed through two message exchanges for the delay-sensitive devices. We also develop Access Pattern Analyzer (APA), which estimates traffic characteristics of MTC devices. When UEs, performing RAPID and contention-based random access, coexist, it is important to determine a value which is the number of preambles for RAPID to reduce random access load. Thus, we analyze random access load using a Markov chain model to obtain the optimal number of preambles for RAPID. Simulation results show RAPID achieves 99.999% reliability with 80.8% shorter uplink latency, and also decreases random access load by 30.5% compared with state-of-the-art techniques.

J. Kim and S. Choi are with the Department of ECE and INMC, Seoul National University, 08826 Seoul, Korea (e-mail: jskim14@mwnl.snu.ac.kr and schoi@snu.ac.kr).

S. Kim is with AI Center in SK Telecom, 04539 Seoul, Korea (e-mail:s1kim@sktair.com)

T. Taleb is with the Department of Communications and Networking, School of Electrical Engineering, Aalto University, 02150 Espoo, Finland, and also with Sejong University, Seoul, Korea (e-mail: tarik.taleb@aalto.fi).

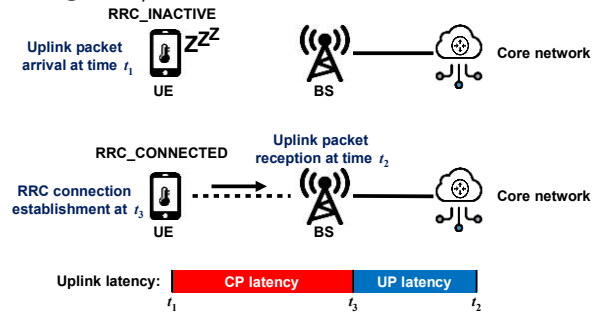

Fig. 1. Uplink packet transmission in RRC INACTIVE state and its latency components.

## I. INTRODUCTION

In recent years, 3GPP mobile communication systems have evolved in a different direction than ever before in order to provide services for Machine Type Communication (MTC) devices [1]. With the development of various services such as eHealth, smart city, and smart factory, the number of MTC connections is expected to grow to 3.3 billion by 2021 [2]. In light of this prediction, 3GPP specifies massive MTC (mMTC) as a new use case of the Fifth Generation (5G) communication systems [3].

MTC devices are battery powered and may be located out of people's reach. Therefore, reducing power consumption of MTC devices is essential to extend their lifetime. In (Long Term Evolution-Advanced) LTE-A system, if a User

This work was supported by Institute of Information & communications Technology Planning & Evaluation (IITP) grant funded by the Korea government (MSIT) (No. 2018-11-1864, Scalable Spectrum Sharing for Beyond 5G Communication). *(Corresponding authors: Seongwon Kim; Sunghyun Choi.)*

Equipment (UE) does not perform any operation for certain amount of time, Base Station (BS) releases connection with the UE, i.e., releases Radio Resource Control (RRC) connection. The connection between BS and core network for the UE is also released. This state is called RRC IDLE state. The UE in RRC IDLE state transits to RRC CONNECTED state to transmit or receive data, by involving a number of message exchanges. However, such transition incurs large random access load for MTC UEs, which frequently transmit small size packets. Therefore, a new RRC state, i.e., RRC _INACTIVE state, is proposed as a primary sleeping state prior to RRC IDLE state [4], [5], and is included in the 3GPP standards [6].

Fig. 1 shows uplink packet transmission when a UE is in RRC INACTIVE state. At time $t_1$, an uplink packet is generated in the UE. Because the UE is in RRC INACTIVE state, random access procedure is needed to establish a RRC connection. If BS receives the packet successfully at time $t_2$, uplink latency[1] is simply computed as $t_2 - t_1$. We can divide uplink latency into two parts. Control Plane (CP) latency is the amount of time to transit from RRC INACTIVE to RRC CONNECTED state. Therefore, $t_3$ is the time when random access procedure is completed. User Plane (UP) latency is the amount of time required for transmitting packets when the UE is active, i.e., in RRC CONNECTED state. The contention-based random access defined in LTE-A increases CP latency as the number of UEs increases. This is because many UEs simultaneously attempt contention-based random access, thus resulting in a preamble collision problem.

---

[1] In this paper, we observe the uplink latency only in the Radio Access Network (RAN), i.e., between UE and BS.

In this paper, we propose RAPID, a novel random access procedure to support delay-sensitive UEs by reducing CP latency. In addition, we develop Access Pattern Analyzer (APA) that works in the RRC layer. APA determines traffic characteristics of MTC UEs for BS to efficiently use radio resources during RAPID operation.

The key idea beneath RAPID is to reduce the number of message exchanges from four to two. One of the components to achieve this purpose is allocating preambles for RAPID by decreasing preambles for contention-based random access.[2] For each procedure, random access load needed for random access increases according to the decrease of the number of preambles for each random access. For this reason, when UEs performing contention-based random access or RAPID coexist, it is important to determine the number of preambles for RAPID. Therefore, we analyze random access load of random access procedures, i.e., contention-based random access and RAPID, using a Markov chain model to determine the optimal number of preambles for RAPID (or preambles for contentionbased random access).

In this paper, we claim the following four major contributions.

- We propose a new random access procedure, RAPID, for delay-sensitive UEs in RRC _INACTIVE state to reduce the uplink latency.
- We develop APA which predicts traffic characteristics of UEs to efficiently use radio resources while UEs perform RAPID procedure.
- Markov chain model is developed to analyze random access load of random access procedures. We also develop an optimization problem to find the optimal number of preambles for RAPID through the analysis.
- We evaluate latency and random access load of RAPID through system-level simulation, and validate that the proposed scheme outperforms state-of-the-art technologies.

The rest of the paper is structured as follows. We discuss the related work in Section II. In Section III, we introduce the RRC states, random access procedure in LTE-A, and uplink latency addressed in this paper. RAPID and APA are detailed in Section IV and Section V, respectively. In Section VI, we analyze random access load of random access procedures, and develop an optimization problem to find the number of preambles for RAPID. We then evaluate the performance of RAPID via simulation under the environment where mMTC UEs exist in Section VII. Finally, the paper concludes in Section VIII.

## II. RELATED WORK

In recent years, many studies have proposed random access procedures for MTC devices [7]. We review two representative random access schemes [8], [9]. We also introduce 2-step random access discussed in 3GPP temporary documents [10], [11]. Lastly, we discuss Sparse Code Multiple Access (SCMA) [12], one of non-orthogonal multiple access schemes.

Prioritized random access: The main idea of this technique is allocating different random access resources for each access class and preventing a large number of UEs from performing random access procedure at the same time. Specifically, it is possible to reduce the competition by allocating different subframe numbers according to each UE's class. Based on this idea, Prioritized Random Access with Dynamic Access barring (PRADA) is proposed [8]. PRADA is proven to be superior to access class barring [13] in terms of random access success probability and average latency. However, since PRADA does not reduce contention among UEs in the same access class, it is difficult to satisfy the latency requirement of delay-sensitive UEs.

Random access for low cost-MTC: 3GPP RAN working group introduced a new Random Access Channel (RACH) structure for Low Cost-MTC (LC-MTC) [9]. RACH for LCMTC consists of multiple narrow band channels. Each channel has a pair of Physical RACH (PRACH) and downlink control channel. The authors of [9] propose a new random access scheme using characteristics of the new RACH structure. In this scheme, using different PRACHs, multiple UEs can transmit the same preamble without collision. Also, BS transmits separate Random Access Responses (RARs) through multiple downlink control channels to reduce the collision of uplink resources. Although this scheme achieves low CP latency by reducing collision probability, it requires four message exchanges which are identical to contention-based random access. Therefore, we reduce the number of massage exchanges from four to two in RAPID to achieve lower CP latency.

2-step random access: In the 3GPP RAN working group, it was agreed that simplified contention-based random access with 2-step should be studied [10], [11]. 2-step random access has the advantage of reducing latency by simplifying the existing contention-based random access procedures. In the first step of 2-step random access, a UE transmits a preamble with payload, i.e., control message or data, using uplink resources randomly selected by UEs. In this case, however, it is possible for the first-step messages from different UEs to collide. Especially, because the probability of collision increases as the number of UEs increases, the 2-step random access proposed in [10], [11] cannot support delay-sensitive UEs. The difference between RAPID and 2-step random access is whether the collision problem can be resolved in two message exchanges. That is, we solve the collision problem in RAPID to achieve lower CP latency.

---

[2] The sum of the number of preambles for RAPID and contention-based random access is fixed.

SCMA: Uplink grant-free transmission based on SCMA allows UEs to transmit data in an arrive-and-go manner. Different UEs may use the same radio resource, but use different codebooks and pilot sequences. In this case, a BS is able to detect the data as long as different codebooks (or pilot sequences) are used [12]. For SCMA operation, uplink synchronization should be maintained with RRC connection established. However, for UEs in RRC INACTIVE state, RRC connection is released and uplink synchronization is also lost. Therefore, SCMA is not appropriate for UEs in RRC INACTIVE state.

## III. PRELIMINARIES

### A. RRC State

In the LTE-A system, only two RRC states are defined, i.e., RRC CONNECTED and RRC _IDLE states. If UE is in RRC _IDLE state, RRC connection must be established to allow the UE to transfer data to BS. Therefore, uplink latency required in RRC IDLE state is larger than in RRC CONNECTED state. BS uses an inactivity timer to manage the RRC state of each UE.

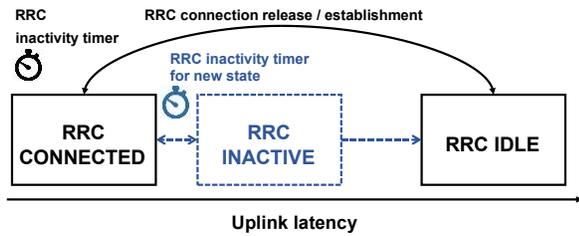

Fig. 2. RRC state machine and state transition in the 5G system: Solid rectangles and arrow represent RRC states and state transition of the LTE-A system, respectively. Dashed rectangle and arrows are newly added in the 5G system.

Fig. 2 shows RRC states and characteristics of each state in the 5G system. In the 5G system, RRC INACTIVE state is introduced [6]. The main characteristics of the new state are as follows. First, while the RRC connection is released, both UE and BS keep the context information of UE's RRC connection, such as UE capabilities and security context. When releasing the RRC connection, the BS allocates an Access Stratum (AS) context ID to the UE in order to activate context information when resuming the RRC connection [14]. Second, connections between the BS and core network for UE remain alive. These properties provide a way to reduce the latency for establishing the RRC connection when UE is in RRC INACTIVE state. Since RRC INACTIVE state is used as the primary sleeping state, the BS should have a new inactivity timer to convert the RRC state from RRC CONNECTED to RRC INACTIVE.

### B. Random Access Procedure

In the LTE-A system, there are two types of random access procedures, i.e., contention-based and contention-free [15]. These procedures can be used for UEs in RRC INACTIVE state.

Contention-based random access: This procedure is initiated by a UE when a BS does not allocate a preamble to the UE. It consists of four steps, and details are as follows.

1) Preamble transmission: The UE transmits a preamble randomly selected from a set of preambles for contentionbased random access. The time when the UE transmits the preamble is determined by a list of allowed time slots allocated by the BS.

2) RAR: The BS, successfully receiving the preamble, transmits RAR including timing advancement value for adjusting the uplink synchronization and uplink resource allocation information for a RRC connection resume request message. The BS and the UE use Random Access-Radio Network Temporary Identifier (RA-RNTI) to transmit and receive RAR, respectively. Specifically, when transmitting (or receiving) RAR, the BS scrambles (or the UE descrambles) bits for error check of control channel with RA-RNTI which is determined by time-frequency resources of preamble transmitted by the UE. BS also includes the received preamble ID in the RAR for UE so that the UE can identify whether the RAR is for itself or not. If the UE does not receive the RAR for certain amount of time, i.e., the size of *RAR window*, the UE tries random access again after performing a backoff procedure.

TABLE I
Latency components of contention-based random access.

| No. | Description | Time (ms) |
|---|---|---|
| 1 | Average delay due to RACH scheduling period | 0.25 |
| 2 | RACH preamble transmission | 0.5 |
| 3 | Preamble detection and RAR transmission | 1.5 |
| 4 | UE processing delay | 1.25 |
| 5 | RRC connection resume request transmission | 0.5 |
| 6 | BS processing delay | 1 |
| 7 | RRC connection setup transmission | 0.5 |
| 8 | UE processing delay | 3 |
| | Total latency | 8.5 |

3) RRC connection resume request: The UE transmits the RRC connection resume request message using uplink resources allocated through the RAR. If two or more UEs simultaneously send the same preamble, UEs transmit RRC connection resume request messages using the same uplink resources. This is because the UEs have the same RARNTI, and thus receive the same RAR. In this case, it is difficult for the BS to successfully decode RRC connection resume request message of each UE.

4) RRC connection setup: If the BS successfully receives the third message, it sends a RRC connection setup message including Cell-RNTI (C-RNTI) to identify the UE in the cell. As UE successfully receives the fourth message, the random access procedure is completed. However, if the UE does not receive the fourth message for certain amount of time, i.e., the value of *contention resolution timer* becomes zero, the UE tries random access again.

Contention-free random access: This procedure is performed when the BS assigns a preamble to the UE transitioning to a RRC INACTIVE state. The preamble is selected from a set of preambles for contention-free random access. The BS that successfully received the preamble transmits the RAR as in the case of contention-based random access. Contention-free random access is completed through exchanging two messages because the preamble does not collide.

### C. Uplink Latency in RRC INACTIVE State

In RRC _INACTIVE state, when an uplink packet is generated, UE must perform random access for resuming the RRC connection. Table I shows contention-based random access procedure and its latency components [16]. We set the Transmit Time Interval (TTI) value to 0.5 ms.[3] In this case, the CP latency becomes 8.5 ms assuming that the processing time is reduced by one-fourth compared with the LTE-A [16]. With this assumption, the UP latency[4] becomes 3 ms. Therefore, the uplink latency is 11.5 ms. If the number of UEs performing contention-based random access increases, the uplink latency even increases further due to preamble collisions, thus increasing the latency requirement that can be satisfied. In the case of contention-free random access, because the procedure is completed in two steps, uplink latency becomes 6.5 ms, i.e., the CP latency is 3.5 ms (No. 1–4 in Table I) and the UP latency is 3 ms. However, it is impossible for a large number of UEs to perform contention-free random access because the number of preambles is required to be equal to the number of UEs.

TABLE II
List of frequently-used parameters.

| Symbol | Description |
|---|---|
| $n(S)$ | Total number of preambles |
| $n(S_{cb})$ | Number of preambles for contention-based random access |
| $n(S_{cr})$ | Number of preambles for RAPID |
| id | AS context ID |
| pid | Preamble ID |
| | UE index |
| | Received uplink packet index |
| | RACH period |
| $T_{ind}$ | Offset index |
| $t_{TTI}$ | TTI value |
| $t_{up}$ | UP latency |
| $t_I$ | Inactivity timer value for transition to RRC INACTIVE state |
| $t_{r-1}$ | Reception time of the $r$-th uplink packet |
| $N_{cb}$ | The number of UEs performing contention-based random access |
| $N_{cr}$ | The number of UEs performing RAPID |
| $N_{ed}$ | The number of UEs whose traffic type is ED among $N_{cr}$ UEs |

## IV. RAPID: PROPOSED RANDOM ACCESS PROCEDURE

### A. Overview

We propose RAPID to overcome the limitations of the conventional random access procedures which are mentioned in Section III-C. The key feature of RAPID is to complete the random access procedure by exchanging only two messages. RAPID enables this by using AS context ID for the following two procedures:
- Selection of the preamble and the set of allowed slot numbers to transmit preamble
- Scrambling of error check bits for control channel when sending RAR

In the proposed scheme, different UEs would try random access by selecting the same preamble in the same slot. However, contention can be resolved by sending different RARs scrambled by AS context ID of each UE. Therefore, RAPID is a contention resolution-based Random Access Procedure using AS context ID. The detailed description of RAPID is provided in the following subsections. Table II provides the list of parameters used in this paper along with their definition.

### B. Criterion of Applying RAPID

The existing random access cannot satisfy the latency requirement of a UE according to the number of UEs served by

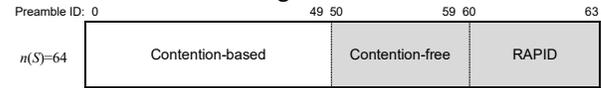

(a) Preamble classification

| Slot number | | | | | | | | | |
|---|---|---|---|---|---|---|---|---|---|
| 0 | 1 | 2 | 3 | 4 | 5 | 6 | 7 | 8 | 9 |

| RACH period ($T_p$)* | Offset index ($T_{ind}$) | Set of allowed slot numbers |
|---|---|---|
| 1 | 1 | 0, 1, 2, 3, 4, 5, 6, 7, 8, 9 |
| 2 | 1 | 0, 2, 4, 6, 8 |
| 2 | 2 | 1, 3, 5, 7, 9 |
| 3 | 1 | 1, 4, 7 |
| 3 | 2 | 2, 5, 8 |
| 3 | 3 | 3, 6, 9 |

*The unit of RACH period is the number of slots

(b) RACH period and offset index

Fig. 3. Preamble set and RACH time resources.

a BS. Therefore, the BS should determine the random access method for the UE in consideration of latency requirement and the number of UEs served by the BS. The BS notifies this information to the UEs via the AS context ID. That is, the AS context ID includes information whether to apply RAPID or not.

---

[3] We assume subcarrier spacing is 30 kHz which is doubled compared with the LTE-A [17]. Therefore, TTI value is 0.5 ms that is halved compared with the LTE-A.

[4] The UP latency value is the time from when UE transmits the Buffer Status Report (BSR) message to when BS successfully receives the data [18].

For example, if the most significant bit of AS context ID[5] is zero, RAPID should be used. Otherwise, the contentionbased random access procedure should be used. We assume that contention-free random access procedure is not used for UEs in RRC INACTIVE states.

*C. Preamble Set and RACH Period Allocation*

A UE selects a preamble and a set of allowed slot numbers to transmit preamble using AS context ID. For this purpose, BS must inform UE of $n(S_{cr})$ representing the number of preambles for RAPID and $T_p$ representing RACH period using a broadcast message. The RACH period is the interval between slot numbers that UE can transmit the preamble. Fig. 3(a) shows an example of preamble allocation to support RAPID. We consider a total $n(S)$ of preambles where $n(S) = 64$ and allocate 10 preambles for contention-free random access [8]. In addition, we allocate four preambles for RAPID. Fig. 3(b) represents sets of allowed slot numbers according to each $T_p$. We consider three types of $T_p$ values. Offset index, denoted by $T_{ind}$, is defined to distinguish the sets of allowed slot numbers when $T_p$ is fixed. For example, when $T_p = 3$, $T_{ind}$ can range from one to three.

*D. Preamble Transmission*

Using AS context ID, UE selects a preamble ID, denoted by pid, from the given preamble set and determines a offset index to transmit the selected preamble.

[5] We assume that AS context ID consists of enough bits to cover the number of UEs we handle.

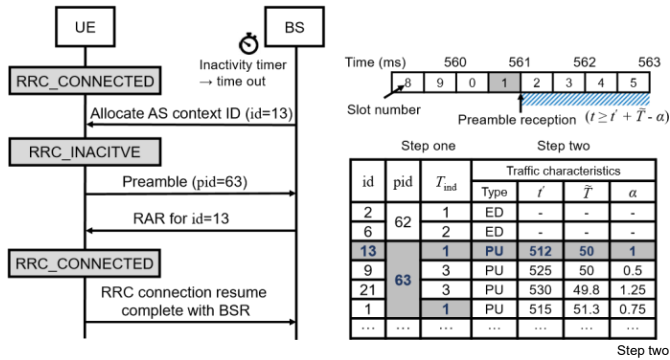

Fig. 4. Overall RAPID procedure of UE whose AS context ID is 13 and process of selecting UEs to receive RAR.

Preamble selection: Using $n(S)$ and $n(S_{cr})$ broadcast by BS, the preamble ID of the $i$-th UE is calculated by

$$\text{pid}(i) = n(S) - 1 - (\text{id}(i) - 1) \mod n(S_{cr}), \quad (1)$$

where $\text{id}(i)$ represents the $i$-th UE's AS context ID in decimal and mod is modulo operation. The zero value is not assigned to $\text{id}(i)$, and each $\text{pid}(i)$ is one of the values from $n(S) - n(S_{cr})$ to 63. Since UE selects the preamble using AS context ID, BS does not need to use additional resources to inform the preamble ID that UE will use in RRC INACTIVE state. Offset index selection: Using $n(S_{cr})$ and $T_p$ broadcast by BS, the offset index of the $i$-th UE is given by

$$T_{\text{ind}}(i) = \left\lfloor \frac{(\text{id}(i) - 1) \mod (n(S_{cr})T_p)}{n(S_{cr})} \right\rfloor + 1. \quad (2)$$

The numerator value of the floor function input can have an integer from 0 to $n(S_{cr})T_p - 1$. If we divide this value by $n(S_{cr})$ and apply the floor function, the output of the function has an integer value from 0 to $T_p - 1$. Therefore, we add one at the end to determine $T_{\text{ind}}(i)$ in (2). The UE transmits the selected preamble in the slot number closest to a slot in which traffic is generated among the set of allowed slot numbers corresponding to $T_{\text{ind}}$.

*E. RAR Transmission*

After receiving preambles for RAPID, BS transmits one or more RARs[5] to candidate UEs that could send the preambles received by BS. When transmitting RARs for each UE, BS scrambles error check bits of control channel for each RAR with id of each candidate UE, respectively. Therefore, UE can successfully receive the RAR by descrambling error check bits of control channel with allocated id. It is important to send RARs to UEs having high probability of sending the preamble. Therefore, the BS should select UEs to receive the RAR through two steps as shown in Fig. 4. In this example, we consider $n(S_{cr}) = 4$ and $T_p = 3$.

1) Step one: The BS has a table that contains allocated AS context ID, preamble ID, offset index, and traffic characteristics. The BS filters UEs that could send the preamble based on the received preamble and the slot number at which the preamble is received. In Fig. 4, for example, the preamble with pid = 63 is received at slot number one, and hence, UEs whose id = 1,13 could send the preamble (Shaded region in Step one columns in Fig. 4).

2) Step two: In this step, the BS additionally exploits the traffic characteristics to predict UEs more precisely. For this purpose, we develop APA to estimate the traffic characteristics, i.e., traffic type, estimated traffic period, and margin value. We consider two traffic types, i.e., Periodic Update (PU) and Event Driven (ED) [19]. The PU traffic continuously generates uplink packets with a constant period, $T$. It should be noted that $T$ is a separate parameter not related to RACH period, i.e., $T_p$. The ED traffic follows a Poisson process traffic model with an arrival rate, $\lambda_{ed}$. In case of the PU traffic, the BS exploits $\tilde{T}$ representing estimated traffic period and $\alpha$ representing margin value obtained from APA. If preamble reception time, denoted by $t$, satisfies the equation below, the BS transmits RAR for that UE (Shaded region in Step two columns in Fig. 4).

[5] Since MTC UEs covered in this paper are fixed in position, timing advancement value obtained when UE first accesses to BS is applied. To this end, the timing advancement value of UE should be stored in AS context.

$$t \geq t^0 + \tilde{T} - \alpha, \quad (3)$$

where $t^0$ is the most recent time at which the BS receives the preamble for a successful random access procedure. In case of ED traffic, the BS always transmits RAR because of the difficulty of predicting traffic characteristics. In Fig. 4, the BS transmits the RAR to UE whose id is 13 because its traffic characteristics satisfy (3). The detailed procedures that APA obtains the traffic characteristics are further described in Section V.

UEs who perform RAPID do not carry out backoff procedure if random access is failed due to channel error or APA operation error. This is because the latency requirement of the delay-sensitive UE should be satisfied using RAPID procedure.

*F. AS Context ID Allocation*

BS allocates AS context ID (id) to UE when its inactivity timer value becomes zero as shown in Fig. 4. Allocating id to a specific UE means that BS determines preamble ID (pid) and offset index ($T_{ind}$) the UE will use. Therefore, the way to allocate id should reflect the following two elements for reducing random access load and satisfying latency requirement.
- Traffic type of UE
- Slot numbers at which BS receives uplink packets when UE is in RRC CONNECTED state

If PU traffic UE and ED traffic UE[6] are allocated the same preamble, random access load increases because BS always sends RAR when receiving preamble allocated to the ED traffic UE. Therefore, BS should allocate different preambles depending on the traffic type of UE. For example, all PU traffic UEs are allocated a common preamble, and each ED traffic UE is randomly allocated one preamble among the remaining $n(S_{cr}) - 1$ preambles.

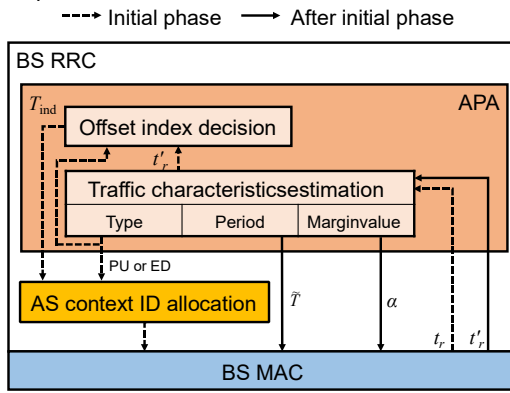

Fig. 5. Overall APA operation.

---

[6] PU (or ED) traffic UE is the UE whose traffic type is the PU (or ED).

In case of PU traffic UE, slot numbers at which BS receives uplink packets are also considered. For instance, when $T_p = 3$, the maximum waiting time for PU traffic UE to transmit preamble is four slots, i.e., 2 ms when $t_{TTI}$ is 0.5 ms. This value could affect the satisfaction of latency requirement for delay-sensitive UEs. Therefore, BS selects $T_{ind}$ based on slot numbers receiving uplink packets when UE is in RRC CONNECTED state. The procedure of determining $T_{ind}$ is detailed in Section V-D. Otherwise, in case of ED traffic UE, because it is difficult to predict traffic characteristics, BS randomly allocates candidate values of $T_{ind}$ to UEs.

In short, when BS receives uplink packets from a UE, BS first determines pid and $T_{ind}$ based on the UE's traffic type and the slot numbers at which the uplink packets are received, respectively. After that, BS randomly selects id among the candidate IDs mapped to the given pid and $T_{ind}$.

*G. Number of Preambles for RAPID*

As the number of preambles for RAPID, i.e., $n(S_{cr})$, increases, fewer ED traffic UEs are allocated to the same pid and $T_{ind}$. Therefore, random access load caused by both unnecessary RAR transmissions and unnecessary uplink resource allocation for RRC connection resume request messages decreases as $n(S_{cr})$ increases. On the other hand, random access load for contention-based random access increases by increasing $n(S_{cr})$. This is because the total number of two types of preambles, i.e., $n(S_{cb}) + n(S_{cr})$, is fixed at 54. For a given scenario, it is therefore of great importance to determine the optimal $n(S_{cr})$ considering such trade-off relationship. For this purpose, in Section VI, we analyze random access load of two random access procedures, i.e., contention-based random access and RAPID, and develop an optimization problem to determine $n(S_{cr})$.

V. ACCESS PATTERN ANALYZER

*A. Overview*

APA predicts traffic characteristics of each UE to help BS allocate AS context ID to UE and send RAR messages during RAPID operation. As mentioned in Section IV-B, the BS determines whether RAPID is applied to each UE in consideration of latency requirement. APA initially estimates

Algorithm 1 APA operation for the $i$-th UE.

Initialize:
1: $t_I \leftarrow T_{init}, r \leftarrow 0$

During initial phase:
2: **while** $t_I \neq 0$ **do**
3:     **if** New uplink packet is received **then**
4:         $r \leftarrow r + 1$
5:         $t_{r-1}(i) \leftarrow$ current time
6:         $t'_{r-1}(i) = t_{r-1}(i) - t_{up} + t_{TTI}$
7:         **if** $r \geq 2$ **then**

```
8:        t(i) = (t'_0(i), · · · , t'_{r-1}(i))^T
9:        t̃'_0(i), T̃_{r-1}(i) ← LR(r, t(i))
10:    end if
11:    if r = R_th then
12:       σ² = Var(T̃_1(i), · · · , T̃_{R_th-1}(i))
13:       if σ² ≤ δ_th then
14:          UE has PU traffic type
15:          k* ← MAS(t(i), R_th)   ▷ Algorithm 2
16:       else
17:          UE has ED traffic type
18:       end if
19:       t_I ← T_I
20:       break
21:    end if
22: else
23:    t_I ← t_I − t_TTI
24: end if
25: end while
26: if r < R_th then
27:    UE has ED traffic type, t_I ← T_I
28: end if
    After initial phase for PU traffic UE:
29: if Preamble for the i-th UE is received then
30:    t_temp(i) ← current time
31:    if Random access succeed then
32:       r ← r + 1
33:       t(i) = (t_0(i), · · · , t_{r-1}(i))
           t̃'_0(i), T̃_{r-1}(i) ← LR(r, t(i))
           t'_{r-1}(i) ← t_temp(i)
34:       t(i) = (t'_0(i), · · · , t'_{r-1}(i))^T
35:
36:    end if
37: end if
```

the traffic type of each delay-sensitive UE, so that we define an initial phase in APA. The dashed arrows in Fig. 5 represent the operation in the initial phase. The traffic type is estimated based on the uplink packet reception time (Section V-B). $T_{ind}$ should be determined to allocate AS context ID. For this purpose, we define offset index decision procedure (Section V-D). After the initial phase, if the traffic type is PU, APA estimates the traffic period and margin value of the corresponding UE (Section V-B and V-C).

### B. APA Operation

During the initial phase, UEs are in RRC _CONNECTED state, and APA obtains reception time values of uplink packets. Algorithm 1 shows the detailed procedure estimating the traffic type. At first, RRC inactivity timer value for transition to RRC INACTIVE state, denoted by $t_I$, is set to $T_{init}$. APA stores the time value of $t_{r-1}(i)$ when BS receives the $r$th uplink packet from the $i$-th UE (line 5). To estimate the time to receive the preamble after the initial phase, we keep track of the $r$-th preamble reception time using $t^0_{r-1}(i)$ which is calculated using $t_{r-1}(i)$ (line 6). Specifically, UP latency, denoted by $t_{up}$, is subtracted from the $t_{r-1}(i)$, and then TTI value, denoted by $t_{TTI}$, is added to indicate the preamble reception time. When more than two uplink packets from the $i$-th UE are received, i.e., $r \geq 2$, the estimated traffic period of the $i$-th UE, denoted by $\tilde{T}_{r-1}(i)$, is calculated by Linear Regression ($LR$) using a normal equation (line 9) [20]. For $r$ uplink packets of the $i$-th UE, the normal equation is defined as

$$\mathbf{X} = \begin{pmatrix} 1 & 0 \\ 1 & 1 \\ \vdots & \vdots \\ 1 & r-1 \end{pmatrix}, \mathbf{t}(i) = \begin{pmatrix} t'_0(i) \\ t'_1(i) \\ \vdots \\ t'_{r-1}(i) \end{pmatrix}$$
$$\Theta = LR(r, \mathbf{t}(i)) = (\mathbf{X}^T \mathbf{X})^{-1} \mathbf{X}^T \mathbf{t}(i), \quad (4)$$

where $(\cdot)^T$ and $(\cdot)^{-1}$ represent transpose and inverse of a matrix, respectively. The result of $LR$, denoted by $\Theta$, is a $2 \times 1$ vector, which is $\left(\tilde{t}'_0(i), \tilde{T}_{r-1}(i)\right)^T$, where $\tilde{t}'_0(i)$ is the estimated initial uplink packet reception time of the $i$-th UE.

When the number of received packets becomes $R_{th}$, the traffic type is estimated based on the variance of the stored estimated traffic period values (lines 11–18). If the UE's estimated traffic type is PU, Algorithm 2 is called to determine $T_{ind}$ of the UE (line 15). After the traffic type of UE is estimated, $t_I$ is set to $T_I$, which is much smaller than $T_{init}$. This is for making the UE go to RRC INACTIVE state. If the traffic type of UE can not be estimated until the timeout of initial inactivity timer, whose value is set to $T_{init}$, the UE is considered having ED traffic type (line 27).[7]

After the initial phase, for the PU traffic UE, APA updates the estimated traffic period and the estimated initial uplink packet reception time (line 35). For this purpose, the time when UE

---

[7] Many MTC applications have periodic traffic with $T$ values much longer than $T_{init}/R_{th}$. $T$ could be in the order of hours, days, or even months. By the way, these applications have latency requirements that can be supported by contention-based random access [21].

receives preamble in the successful RAPID procedure is stored to $t^0_{r-1}$ (line 33).

### C. Margin Value

Accurate period estimation is difficult due to channel errors. Therefore, a margin value is required and it is given by

$$\alpha_{R-1} = \frac{1}{R}\sum_{r=1}^{R}\left|t'_{r-1} - \left(\tilde{t}'_0 + (r-1)\tilde{T}_{R-1}\right)\right|, \quad R \geq 2, \quad (5)$$

where $\tilde{t}'_0$ is the output of (4) and $R$ is the number of $t'_{r-1}$ samples. If we have a total of $R$ samples of $t'_{r-1}$, generalization of (3) is given by

$$t \geq t^0_{R-1} + \tilde{T}_{R-1} - \alpha_{R-1}, \quad R \geq 2. \quad (6)$$

**Algorithm 2** MAS: procedure to determine $k^*$.

Input:
1: **t**, $R_{th}$
Initialize:
2: $s_k \leftarrow 0$ for $\forall k$ ($1 \leq k \leq T_p$)
Call by Algorithm 1
3: switch $T_p$ do
4:     case 1 break
5:     end case
6:     case 2
7:         for $r \leftarrow 1 \ldots R_{th}$ do
8:             $k \leftarrow (t'_{r-1}/t_{\text{TTI}} - 1) \bmod T_p + 1$
9:             $s_k \leftarrow s_k + 1$
10:        end for
11:    end case
12:    case 3
13:       for $r \leftarrow 1 \ldots R_{th}$ do
14:         temp $\leftarrow t'_{r-1}/t_{\text{TTI}} - 1$
15:         if temp = 0 then
16:            $k \leftarrow 1$
17:         else
18:            $k \leftarrow temp \bmod T_p$
19:           if $k = 0$ then
20:                $k \leftarrow 3$
21:           end if
22:         end if
23:         $s_k \leftarrow s_k + 1$
24:       end for
25:    end case
26: end switch
27: $k^* = \underset{k}{\operatorname{argmax}}\, s_k$

### D. Offset Index Decision

As mentioned in Section IV-F, BS determines a UE's pid using the traffic type obtained from APA. For ED traffic UEs, BS randomly selects one of candidate values of $T_{\text{ind}}$. For PU traffic UEs, BS should determine $T_{\text{ind}}$ considering slot numbers receiving uplink packets. For this purpose, BS needs to know a set of allowed slot numbers in which UE will transmit preamble. We refer to this set of allowed slot numbers as the most accessed set, denoted by $s_{k^*}$.

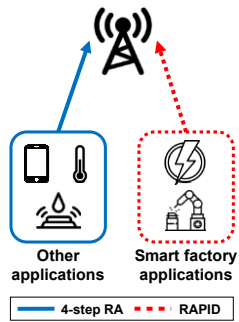
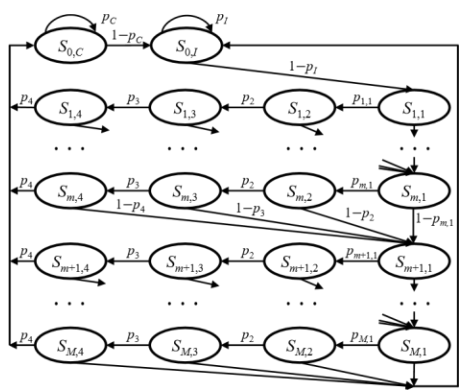
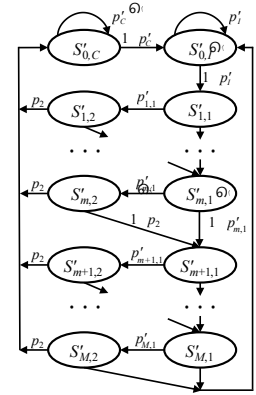

(a) System model      (b) Model for 4-step RA      (c) Model for RAPID

Fig. 6. System model and Markov chain models for random access procedures.

Algorithm 2 shows how to determine $k^*$ using **t** in (4). Firstly, we define $s_k$ (where $1 \leq k \leq T_p$) to investigate frequency of each set (line 2). In case of $T_p = 1$, because there is only one offset index, $k^*$ is one (line 4). If $T_p = 2$ or 3, for all the values of $t'_{r-1}$ in **t**, we can obtain value $k$, which is the index of the set containing the received slot number (lines 6–25). Especially, when $T_p = 3$, if the received slot number is zero, i.e., *temp* is zero, $k = 1$ is appropriate to make waiting time minimum (line 16). Lastly, $k^*$ is the value of $k$ with the largest value $s_k$ (line 27). We name this algorithm Most Accessed Set (MAS), which is a procedure to find $T_{\text{ind}} = k^*$. In addition, a proper $R_{th}$ value should be chosen to obtain $k^*$ according to the value of $T_p$, e.g., when $T_p = 2$, $R_{th}$ should be an odd number.

## VI. RANDOM ACCESS LOAD ANALYSIS

We analyze random access load of two types of random access, i.e., contention-based random access (4-step RA) and RAPID as presented in Section IV-G. We define random access load as the number of scheduled signals including both the untransmitted signals after unnecessary resource allocation and the transmitted signals. We also develop an optimization problem to determine the number of preambles for RAPID, i.e., $n(S_{cr})$, using the analysis result.

### A. System Model

We consider a scenario whereby many MTC UEs in RRC INACTIVE state transmit uplink packets. Each UE has one application with one of two types of latency requirements, i.e., delay-sensitive or delay-tolerant [22], [23]. As shown in Fig. 6(a), UEs running smart factory applications are delay-sensitive devices [24]. The UEs with smart factory applications should use RAPID to meet the latency requirement. The latency requirements of other applications are delay-tolerant such that UEs with those applications perform 4-step RA. The UEs performing 4-step RA should succeed random access procedures within the maximum number of random access opportunities. Under this scenario, we analyze random access load of each random access procedure to find the optimal number of preambles for RAPID, i.e., $n(S_{cr})$. We employ Markov chain to model each random access procedure. In the proposed model, the following assumptions are made.

- We assume packet arrival process follows Poisson [25].
- In 4-step RA, physical random access channel is available in every slot [25].
- In 4-step RA, collision of control messages, i.e., RRC connection request, occurs with constant and independent probability [26].
- Radio resources used for random access procedures are enough to serve the UEs trying each random access.[8]

### B. Markov Chain Model for 4-Step RA

Fig. 6(b) shows the Markov chain model for 4-step RA to analyze random access load, where there are three types of states, i.e., RRC CONNECTED, RRC _INACTIVE, and random access procedure states. The state $S_{0,C}$ represents RRC CONNECTED state. A UE in this state transmits uplink packets and a BS operates inactivity timer of which value is denoted by $t_I$ for that UE. The state $S_{0,I}$ represents RRC INACTIVE state, where the UE waits for the generation of uplink packet. Rest of the states, $S_{m,n}$'s (where $1 \leq m \leq M$ and $1 \leq n \leq 4$), represent 4-step RA procedure. Index $m$ is the number of random access attempts, and $M$ is the maximum number of random access opportunities. Index $n$ represents each step of 4-step RA presented in Section III-B.

For a UE in the state $S_{0,C}$, if an uplink packet is generated before the timer value $t_I$ becomes zero, the UE stays in that state. On the other hand, if nothing happens until $t_I$ becomes zero, a state transition takes place from $S_{0,C}$ to $S_{0,I}$. For the UE in the state $S_{0,I}$, if an uplink packet is generated, the state is transferred from $S_{0,I}$ to $S_{1,1}$. Since the packet arrival process follows a Poisson process, inter-packet arrival time follows an exponential distribution, denoted by $X \sim \mathrm{Exp}(\lambda_{cb})$, where $\lambda_{cb}$ is the packet arrival rate of the UE performing 4-step RA. We can obtain transition probability values for $S_{0,C}$ and $S_{0,I}$ as $p_C = 1 - e^{-\lambda_{cb}(t_{up}+t_I)}$ and $p_I = e^{-\lambda_{cb} t_{TTI}}$, respectively. In the $m$-th random access trial, if the UE fails to transmit (or receive) message for 4-step RA, the state is transferred from $S_{m,n}$ to $S_{m+1,1}$. When $m = M$, if the UE fails, the next state is $S_{0,I}$.

We define transition probability in each step of 4-step RA as follows.

- $p_{m,1}$: This value means preamble detection probability, i.e., $1 - e^{-m}$, where $m$ indicates the $m$-th preamble transmission [25], [27]. Even if multiple UEs transmit the same preamble and collision occurs, the BS can detect the preamble if at least one preamble transmission succeeds without channel error.
- $p_2$ and $p_4$: These values represent successful downlink control messages reception ratio. For these messages, because BS uses very low modulation and coding scheme with high transmission power, we consider $p_2 = p_4 \approx 1$ [25].[9]
- $p_3$: This value means successful transmission probability of RRC connection resume request message. $p_3 = (1 - \rho_{col})(1 - \rho_{ch})$, where $\rho_{ch}$ is channel error, can be reduced to $1 - \rho_{col}$ because of $1 - \rho_{ch} \approx 1$ like $p_2$ and $p_4$.

The collision probability, denoted by $\rho_{col}$, is determined as

$$\rho_{col} = \sum_{j=1}^{N_{cb}-1} \binom{N_{cb}-1}{j} \tau^j (1-\tau)^{N_{cb}-1-j} \times \left(1 - \left(1 - \frac{1}{n(S_{cb})}\right)^j\right), \quad (7)$$

where $N_{cb}$ is the number of UEs performing 4-step RA and $\tau$ means the probability that a UE successfully transmits a preamble in one slot. Eq. (7) represents the probability that when one UE selects a preamble, at least one of the other UEs successfully transmit the same preamble in the same slot.

We denote $\pi_{m,n}$ as stationary probability of state $S_{m,n}$. First, $\pi_{0,I}$ is obtained as

$$\pi_{0,I} = \frac{(1-p_C)\pi_{0,C} + \pi_{M,1}(1-p_{M,1})}{1-p_I} + \frac{\sum_{n=2}^{4} \pi_{M,n}(1-p_n)}{1-p_I}. \quad (8)$$

We can simply calculate $\pi_{1,1} = \pi_{0,I}(1-p_I)$. Next, we define $f(m)$ and $g(m)$ to represent stationary probabilities of $\pi_{m,1}$ and $\pi_{m,n}$ (where $2 \leq n \leq 4$), respectively.

---

[8] In LTE system, downlink control channel resources for random access are limited [27]. For MTC application, however, enough resources for downlink control channel should be guaranteed to improve the random access performance [9], [28].

[9] We validate this value is appropriate by comparing simulation using channel model [29] in Section VI-F.

$$f(m) = \pi_{m,1}$$

$$= \begin{cases} \pi_{0,I}(1-p_I), & m = 1 \\ f(m-1)\Big(1 - p_{m-1,1} + p_{m-1,1} \\ \times \big(1 - p_2 + \sum_{l=3}^{4}\prod_{j=2}^{l-1} p_j(1-p_l)\big)\Big), & m \geq 2, \end{cases}$$

$$g(m) = \sum_{n=2}^{4} \pi_{m,n}$$

$$= p_{m,1}f(m) + p_{m,1}\sum_{n=2}^{3}\prod_{j=2}^{n} p_j f(m)$$

$$= p_{m,1}f(m)\left(1 + \sum_{n=2}^{3}\prod_{j=2}^{n} p_j\right). \tag{9}$$

Because the sum of stationary probabilities of all states is one, we can obtain the following equation using (9).

$$\pi_{0,C} + \pi_{0,I} + \sum_{m=1}^{M}\sum_{n=1}^{4} \pi_{m,n} = 1,$$

$$\pi_{0,C} + \pi_{0,I} + \sum_{m=1}^{M}(f(m) + g(m)) = 1,$$

$$\pi_{0,C} + \pi_{0,I} + \sum_{m=1}^{M} f(m)\left(1 + p_{m,1}\left(1 + \sum_{n=2}^{3}\prod_{j=2}^{n} p_j\right)\right) = 1. \tag{10}$$

Eq. (10) is the function of $\rho_{\text{col}}$ so that $\pi_{0,C}$ can be derived in terms of $\rho_{\text{col}}$. The other stationary probabilities of all states are also derived in terms of $\rho_{\text{col}}$. Meanwhile, $\tau$ also represents the proportion of time in successful preamble transmission of the states $S_{m,1}$ (where $1 \leq m \leq M$). Therefore, $\tau$ is obtained as

$$\tau = \frac{1}{T_{\text{tot}}}\sum_{m=1}^{M}\pi_{m,1}t_{TTI}p_{m,1}, \tag{11}$$

where $T_{\text{tot}}$ is the average holding time for all states, i.e.,

$$T_{\text{tot}} = \pi_{0,C}T_{0,C} + \pi_{0,I}T_{0,I} + \sum_{m=1}^{M}\sum_{n=1}^{4}\pi_{m,n}T_{m,n}, \tag{12}$$

where $T_{m,n}$ is the holding time for each state $S_{m,n}$. $T_{0,C}$ is calculated by [30, Eq. (9)]

$$T_{0,C} = E\big[\min(X, t_{\text{up}} + t_I)\big]$$
$$= \int_0^{\infty} P\big(\min(X, t_{\text{up}} + t_I) > x\big)dx$$
$$= \int_0^{t_{\text{up}}+t_I} P(X > x)dx = \int_0^{t_{\text{up}}+t_I} e^{-\lambda_{\text{cb}}x}dx$$
$$= \frac{1}{\lambda_{\text{cb}}}\left(1 - e^{-\lambda_{\text{cb}}(t_{\text{up}}+t_I)}\right), \tag{13}$$

where $X$ is a random variable representing inter-packet arrival time. Because preamble transmission is possible in every slot, $T_{0,I}$ is $t_{TTI}$, which is the slot length.

The remaining state holding time values can be calculated using Table I. Each state holding time consists of two components, i.e., when each step (index $n$) of the random access procedure succeeds ($p_n$) or fails ($1-p_n$). Therefore, the states $S_{m,n}$ with same $n$ have the same holding time value. The holding time of the state $S_{m,1}$ is

$$T_{m,1} = t_{TTI}p_{m,1} + (t_{TTI} + W_{RAR} + BW_{\text{avg}})(1 - p_{m,1}), \tag{14}$$

where $W_{RAR}$ is the *RAR window* size, and $BW_{\text{avg}}$ is the average value of backoff window size. The holding time of the state $S_{m,2}$ is written as

$$T_{m,2} = 1.5p_2 + (W_{RAR} + BW_{\text{avg}})(1 - p_2). \tag{15}$$

The value 1.5 is the time value of preamble detection and RAR transmission (No. 3 in Table I). The holding time of the state $S_{m,3}$ is obtained as

$$T_{m,3} = 1.75p_3 + (1.75 + W_{\text{res}} + BW_{\text{avg}})(1 - p_3), \tag{16}$$

where $W_{\text{res}}$ is the *contention resolution timer* value. The value 1.75 is the sum of UE processing delay and RRC message transmission time (No. 4–5 in Table I). Lastly, the holding time of the state $S_{m,4}$ is

$$T_{m,4} = 4.5p_4 + (W_{\text{res}} + BW_{\text{avg}})(1 - p_4). \tag{17}$$

The value 4.5 contains BS processing delay, RRC message transmission time, and UE processing delay (No. 6–8 in Table I).

The value $\tau$ can be obtained by solving system of equations with unknown variables $\rho_{\text{col}}$ and $\tau$, i.e., (7) and (11). Specifically, the right-hand side of (11) is changed to the formula in terms of $\tau$. Then, we select the intersection point with $y = \tau$, i.e., left-hand side of (11), to obtain $\tau$. Finally, we can calculate $\rho_{\text{col}}$ and all stationary probabilities using $\tau$.

*C. Average Random Access Load for 4-Step RA*

In 4-step RA, random access load for each state transition is one except for four transitions, i.e., transitions from $S_{0,C}$ and $S_{0,I}$. For each state $S_{m,n}$, random access load is $1 \times \pi_{m,n}p_n + 1 \times \pi_{m,n}(1-p_n) = \pi_{m,n}$. (When $n$ is one, $p_n$ is replaced by $p_{m,1}$.) Therefore, the average random access load for 4-step RA is obtained as

$$E[L_{\text{cb}}] = \frac{1}{T_{\text{tot}}}\sum_{m=1}^{M}\sum_{n=1}^{4}\pi_{m,n}. \tag{18}$$

*D. Markov Chain Model for RAPID*

Fig. 6(c) shows the Markov chain model for RAPID. We only investigate random access load for ED traffic UEs. This is because one of $n(S_{\text{cr}})$ is always allocated to PU traffic UEs, while $n(S_{\text{cr}})-1$ preambles are used for ED traffic UEs as mentioned in Section IV-G. Thus, changing $n(S_{\text{cr}})$ only affects the random access load for ED traffic UEs. The model for RAPID also has three types of states like the model for 4step RA. Because RAPID procedure is completed within two steps, the difference from the model for 4-step RA is that the maximum value of $n$ is two.

The inter-packet arrival time of ED traffic also follows an exponential distribution, i.e., $X^0 \sim \text{Exp}(\lambda_{ed})$, where $\lambda_{ed}$ is packet arrival rate of the ED traffic UE. Therefore, the value of $p_{0C}$ is $1 - e^{-\lambda_{ed}(t_{up}+t_i)}$ and $p'_I$ is $e^{-\lambda_{ed} T'_p t_{TTI}}$. It should be noted that, in the exponent of $p'_I$, we use $T'_p$ instead of $T_p$. $T'_p$ is introduced to represent the average interval of state transitions from $S_{0,I}$ and given as[10]

$$T'_p = \begin{cases} T_p, & T_p = 1,2 \\ 10/3, & T_p = 3. \end{cases} \quad (19)$$

The value of transition probability, $p_2$, is the same as the value for 4-step RA, but $p'_{m,1}$ is different from the value of $p_{m,1}$. In RAPID, we consider the number of UEs allocated to the same offset index and preamble. When $k$ UEs transmit the same preamble in the same slot, if at least one preamble transmission is successful, the BS can transmit one or more RARs. Therefore, in RAPID, preamble detection probability for a UE transmitting the $m$-th preamble is obtained as

$$p'_{m,1} = 1 - \prod_{j=1}^{M} \left(e^{-j}\right)^{\overline{N}'_{UE}(j)},$$

$$\overline{N}'_{UE}(j) = \begin{cases} \overline{N}_{UE}(j), & j = m, \\ \overline{N}_{UE}(j) - 1, & j \neq m, \end{cases} \quad (20)$$

where $\overline{N}_{UE}(j)$ is the average number of UEs transmitting the $j$-th preamble at the same time. The value $\overline{N}_{UE}(j)$ is

$$\overline{N}_{UE}(j) = \sum_{k=1}^{\lceil \overline{N}_{RAR} \rceil} p_k(j) k, \quad (21)$$

where $p_k(j)$ is the probability that $k$ UEs transmit the same preamble simultaneously, i.e., transmit the $j$-th preamble transmission at the same time, when one UE transmits a preamble in a particular slot.

$$p_k(j) = \binom{\lceil \overline{N}_{RAR} \rceil - 1}{k - 1} \left(\frac{T'_p}{\text{slot}_{avg}}(1 - p_{j-1,1})\right)^{k-1} \times \left(1 - \frac{T'_p}{\text{slot}_{avg}}(1 - p_{j-1,1})\right)^{\lceil \overline{N}_{RAR} \rceil - k}, \quad (22)$$

where $\overline{N}_{RAR}$ is the average number of ED traffic UEs uniformly allocated to the combinations of offset indices and the number of preambles that ED traffic UEs use, which is

$$\overline{N}_{RAR} = \max\left(\frac{N_{ed}}{T_p(n(S_{cr}) - 1)}, 1\right). \quad (23)$$

$\text{slot}_{avg}$ is the average number of slots for which one packet is generated for a UE, so that $\text{slot}_{avg} = 1/(\lambda_{ed} t_{TTI})$.

$\pi'_{m,n}$ is the stationary probability of state $S'_{m,n}$. As we did in Section VI-B, $\pi'_{0,I}$ is obtained as

$$\pi'_{0,I} = \frac{(1 - p'_c)\pi'_{0,C} + \pi'_{M,1}(1 - p'_{M,1}) + \pi'_{M,2}(1 - p_2)}{1 - p'_I}, \quad (24)$$

and $\pi'_{1,1}$ is $\pi'_{0,I}(1 - p'_I)$. We also define $f'(m)$ to represent the stationary probability of state $S_{m,1}$.

$$f'(m) = \begin{cases} \pi'_{0,I}(1 - p'_I), & m = 1 \\ f'(m-1)(1 - p'_{m-1,1} \\ \quad + p'_{m-1,1}(1 - p_2)), & m \geq 2. \end{cases} \quad (25)$$

We also derive the following equation using the property that the sum of all stationary probabilities is one.

$$\pi'_{0,C} + \pi'_{0,I} + \sum_{m=1}^{M} \sum_{n=1}^{2} \pi'_{m,n} = 1,$$

$$\pi'_{0,C} + \pi'_{0,I} + \sum_{m=1}^{M} (1 + p'_{m,1}) f'(m) = 1 \quad (26)$$

Because there is no unknown variable in (26), we can obtain $\pi'_{0,C}$. Also, the other stationary probabilities are calculated.

*E. Average Random Access Load for* RAPID

The average random access load for RAPID can be obtained as

$$E[L_{ed}] = \frac{1}{T'_{tot}} \sum_{m=1}^{M} \pi'_{m,1} + (2\text{eff}(N_{RAR}) - 1)\pi'_{m,2}, \quad (27)$$

where $T'_{tot}$ is the average holding time for all states, which is

$$T'_{tot} = \pi'_{0,C} T'_{0,C} + \pi'_{0,I} T'_{0,I} + \sum_{m=1}^{M} \sum_{n=1}^{2} \pi'_{m,n} T'_{m,n}. \quad (28)$$

---

[10] For the state $S_{0,I}$, a state transition occurs at the allowed slots. In RAPID, however, the number of slots between two consecutive allowed slots is not always equal to $T_p$. For instance, if the packet is generated in slot number seven when $T_p = 3$ and $T_{ind} = 1$, preamble is transmitted in the next slot number one.

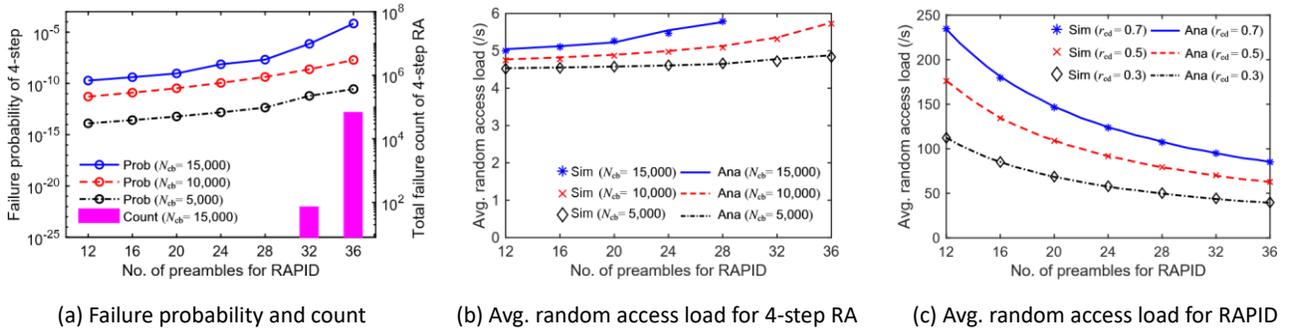

(a) Failure probability and count    (b) Avg. random access load for 4-step RA    (c) Avg. random access load for RAPID

Fig. 7. The left-side graph represents the failure probability of 4-step RA for UE in the left axis, and the total failure count of 4-step RA in the right axis ('Prob' and 'Count' represent failure probability and total failure count, respectively). The two right-side graphs show the average random access load for 4-step RA and RAPID ('Sim' and 'Ana' represent simulation and analysis results, respectively).

$T'_{0,C}$ is calculated in the same way as $T_{0,C}$. In the result of (13), only $\lambda_{cb}$ is changed to $\lambda_{ed}$. The holding time for the state $S'_{0,I}$ is $T'_{0,I} = t_{\text{TTI}} T'_p$. The values of holding time for states $S_{m,1}$ and $S'_{m,2}$ $m,0$ are $T_{m,0\ 1} = t_{\text{TTI}} p\ 1 + (t_{\text{TTI}} + W_{\text{RAR}})(1 - p_{m,1})$ and $T'_{m,2} = 2.75 p_2 + W_{\text{RAR}}(1 - p_2)$, respectively.

For the random access load for state transition from $S_{m,0\ 1}$ is one, and for state transition from $S_{m,0\ 2}$, we can exploit the result of (23), which is $\overline{N}_{\text{RAR}}$. It also represents the average number of RAR messages a BS transmits after receiving a preamble. However, it is not accurate to use the value of $\overline{N}_{\text{RAR}}$ to obtain the average random access load. This is because when $k$ UEs transmit the same preamble in the same slot, random access load due to multiple RAR transmissions is reduced from one UE's perspective. Therefore, we calculate effective random access load, denoted by $\text{eff}(\overline{N}_{\text{RAR}})$, reflecting the probability that $k$ UEs transmit the same preamble in the same slot, and it is calculated by

$$\text{eff}(\overline{N}_{\text{RAR}}) = \frac{\sum_{k=1}^{\overline{N}_{\text{RAR}}} p'_k \frac{\text{RAR}}{k} d}{\overline{N}_{\text{RAR}}}, \quad (29)$$

where $p'_k$ is the probability that one or more UEs transmit the same preamble simultaneously when one UE transmits a preamble in a particular slot, i.e.,

$$p_{0k} = \frac{\binom{\lceil \overline{N}_{\text{RAR}} \rceil - 1}{k-1} \left( \frac{T'_p}{\text{slot}_{\text{avg}}} \sum_{m=0}^{M-1}(1-p_{m,1}) \right)^{k-1}}{\left(1 - \frac{T'_p}{\text{slot}_{\text{avg}}} \sum_{m=0}^{M-1}(1-p_{m,1})\right)^{d\overline{N}_{\text{RAR}}-k}} \times \quad (30)$$

When the average random access load of RAPID is calculated, $\text{eff}(\overline{N}_{\text{RAR}})$ is doubled and then one is subtracted as in (27). This means that random access load includes not only all RAR transmissions but also unnecessary uplink resource allocation for RRC connection resume request messages except uplink resource allocation for a UE who transmits a preamble, i.e., this allocation is used for the transmission of an RRC connection resume request message.

### F. Validation of Analysis

For the validation of analysis, we introduce scenarios that consist of two types of applications defined in Section VI-A.

TABLE III
System parameters.

| Parameter | Value | Parameter | Value |
|---|---|---|---|
| Carrier frequency | 2 GHz | Avg. backoff window size | 5 ms |
| System bandwidth | 20 MHz | Contention resolution timer | 24 ms |
| $t_{\text{TTI}}$ | 0.5 ms | $T_{\text{init}}$ | 5 s |
| $n(S_{cr}) + n(S_{cb})$ | 54 | $T_I$ | 5 ms |
| RACH period ($T_p$) | 3 slots | $R_{\text{th}}$ | 10 |
| RAR window size | 2.5 ms | $\delta_{\text{th}}$ | 0.1 |

Table III summarizes system parameters for analysis and simulation [31], [32]. Specifically, the parameters related to random access are revised considering the reduced $t_{\text{TTI}}$, i.e., 0.5 ms. For the simulation, we create path loss and shadowing following 3D-Urban Macro (3D-UMa) model defined in [33]. We also consider fast fading channel model generated using ITU-R IMT UMa model in [29].

Since we have to find an appropriate number of preambles for RAPID, i.e., $n(S_{cr})$, we first determine the upper bound of $n(S_{cr})$. This is the same way that we find the minimum number of preambles for 4-step RA to guarantee the reliability of UEs performing 4-step RA. For this purpose, we observe two values: (i) the failure probability of 4-step RA for a UE in the analysis and (ii) the total failure count of 4-step RA in the simulation. The failure probability of 4-step RA is given by

$$P_{\text{fail,cb}} = \pi_{M,1}(1 - p_{m,1}) + \sum_{n=2}^{4} \pi_{M,n}(1 - p_n), \quad (31)$$

where $M$ is the maximum number of random access attempts, which is 10 [21]. The total failure count of 4-step RA is the number of failures in all of 10 4-step RA attempts. Fig. 7(a) shows the above two values under the various $n(S_{cr})$ and the number of UEs performing 4-step RA, i.e., $N_{cb}$. Traffic arrival rate, i.e., $\lambda_{cb}$, is fixed at 1 packet/s. When $N_{cb}$ is 5,000 or 10,000, the maximum value of total failure count is under ten, so that we do not present it in Fig. 7(a). When $N_{cb}$ and $n(S_{cr})$ are 15,000 and 32, respectively, the value of total failure count starts increasing. That is why, in certain scenarios, if the number of preambles in 4-step RA, i.e., $n(S_{cb})$, is smaller than a certain value, the number of UEs to try random access continues to increase due to collision, and random access repeatedly fails. Therefore, we determine a threshold value of the failure probability as $10^{-7}$ to prevent the total failure count from starting to increase. We define the reliability of 4-step RA, which is $1 - P_{\text{fail,cb}}$. Therefore, we only consider $n(S_{cr})$ up to the value satisfying the reliability of 4-step RA above 99.99999%.

Fig. 7(b) shows the average random access load of 4-step RA for a UE according to $n(S_{cr})$. When the number of UEs performing 4-step RA is fixed, we can observe the average random access load slightly increases as $n(S_{cr})$ increases. Fig. 7(c) shows the average random access load of RAPID for an ED traffic UE according to $n(S_{cr})$. The number of UEs performing RAPID, denoted by $N_{cr}$, is fixed at 1,000. We consider various ratios of the number of ED traffic UEs to the total number of UEs, which is $r_{ed} = 0.3, 0.5, 0.7$. Traffic arrival rate of ED traffic UE, i.e., $\lambda_{ed}$, is 6.8 packet/s [24]. The random access load is reduced as $r_{ed}$ decreases and $n(S_{cr})$ increases, because fewer ED traffic UEs are allocated to the same combination of preamble and allowed slot numbers.

### G. Optimization Problem

We develop the optimization problem for determining $n(S_{cr})$ to minimize the sum of random access loads of 4step RA and RAPID for ED traffic UEs. The optimal number of preambles for RAPID is argmin $E[L_{cb}]N_{cb} + E[L_{ed}]N_{ed}$
$n(S_{cr})$

$$\text{subject to } n(S_{cr}) = j \ (0 \leq j \leq 54) \quad (32)$$
$$P_{\text{fail,cb}} < 10^{-7},$$

where 10 preambles are assigned for contention-free random access. Therefore, $n(S_{cr})$ can have the value which is from zero to 54. Also, as observed in Section VI-F, $P_{\text{fail,cb}}$ is less than $10^{-7}$ to satisfy the reliability of 4-step RA above 99.99999%.

Since we cannot represent the objective function by a closed form, the above optimization problem should be solved using an exhaustive search. Accordingly, we calculate the values of the objective function depending on $n(S_{cr})$, i.e., zero to 54. That is, the value of the objective function should be calculated up to 55 times. Objective function consists of two terms. The second term which is (27) can be easily obtained from solving (26), which is a linear equation. The first term can be a bottleneck in terms of computational complexity. As mentioned in Section VI-B, the system of equations composed of (7) and (11) should be solved. At this time, it can be a high-order equation depending on $N_{cb}$. We can obtain computational complexity as $O(n^3)$ by using subdivision algorithm to compute isolating intervals for the real roots of a $n$-th order polynomial [34], [35].

The value of the first term in the objective function increases as the number of preambles for RAPID increases because of increase of the collision probability. On the other hand, the value of the second term in the objective function decreases as the number of preambles for RAPID increases because of decrease of the unnecessary RAR transmission. Also, the value of the second term increases as $r_{ed}$ increases because of increase of unnecessary RAR transmission. The optimal number of preambles for RAPID is the value of $n(S_{cr})$ that yields the minimum value of the objective function.

TABLE IV
Simulation scenario.

| Applications | Smart factory | | Other types |
|---|---|---|---|
| Traffic type | PU | ED | ED |
| $T$ or $\lambda$ | 50 ms | 6.8/s | 0.5/s |
| The number of UEs | 1,000 | | 23,000 |
| $r_{ed}$ | 0.3, 0.5, 0.7 | | - |

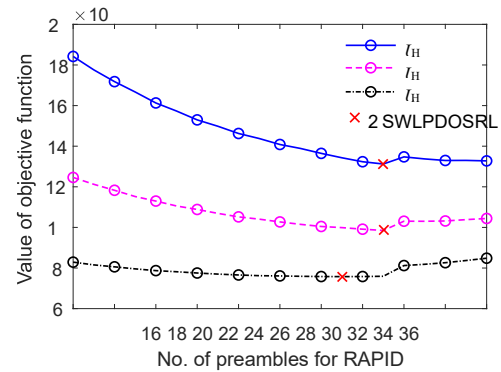

Fig. 8. The optimal number of preambles for RAPID under the scenario in Table IV.

## VII. PERFORMANCE EVALUATION

In this section, we evaluate the performance of RAPID via MATLAB simulation. 4-step RA, PRADA [8], and LC-MTC Random Access (LC-MTC RA) [9] presented in Section II are used as comparison schemes.

### A. Simulation Setup

Table IV shows the parameters of the scenario in order to evaluate RAPID. The total number of UEs connected with a BS is 24,000 [31], and the number of UEs with smart factory application is 1,000. For smart factory application, period of PU traffic UEs is 50 ms and traffic arrival rate of ED traffic UEs is 6.8 packets/s [24]. Traffic arrival rate of UEs with the other applications is 0.5 packet/s. The packet sizes of PU traffic and ED traffic are 125 bytes and 10 bytes, respectively [31]. We use the system parameters defined in Table III.

*B. Number of Preambles for* RAPID

We now provide the optimal value of $n(S_{cr})$ for each scenario in Table IV by solving the optimization problem in (32) using MATLAB. Fig. 8 shows the value of objective function according to $n(S_{cr})$ under various ratio of ED traffic UE. First, as mentioned in Section VI-F, we determine the minimum number of preambles for 4-step RA ensuring reliability above 99.99999%. In case of our simulation scenario, the minimum number of preambles for 4-step RA is 18. Thus, we observe the value of $n(S_{cr})$ until 36. The appropriate $n(S_{cr})$ minimizes the random access load while ensuring the reliability of 4-step RA. In the rest of simulation for RAPID, therefore, we set the values of $n(S_{cr})$ to 29, 31, and 31 when $r_{ed}$ is 0.3, 0.5, and 0.7, respectively.

*C. Performance of* RAPID

Latency and Reliability: In general, reliability is defined as the probability that a certain amount of data from a user device will be successfully transmitted to another peer within a predetermined time [36]. The predetermined time is uplink latency requirement value, which is denoted by $L_{rq}$. Accordingly, the reliability can be expressed as

$$\text{Reliability} = P(D \leq L_{rq}), \quad (33)$$

where $D$ is the measured uplink latency. As the reliability is a function of the uplink latency requirement, we can obtain the corresponding latency requirement value for a given reliability. We define the value as satisfiable latency requirement with the given reliability.

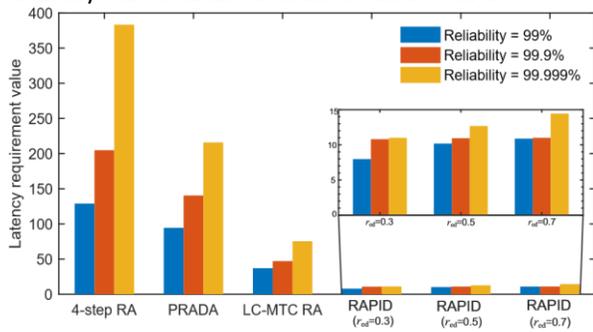

Fig. 9. Satisfiable latency requirements with various reliability values of PU traffic UEs.

Fig. 9 shows satisfiable latency requirements with different reliability values of PU traffic UEs for each scheme. In 4-step RA, we can find the latency requirement with 99% reliability is larger than 100 ms. In this scenario, it is hard to satisfy the latency requirement of delay-sensitive UEs by 4-step RA. In the case of PRADA and LC-MTC RA, the satisfiable latency requirement increases sharply as the reliability value to be satisfied increases. In RAPID, when $r_{ed}$ is 0.7, the satisfiable latency requirement with 99.999% reliability is 14.44 ms, which is 80.8% smaller than that of LC-MTC RA. The reason for the latency decrease in RAPID is the higher random access success probability in virtue of multiple RAR transmissions after a successful preamble transmission from at least one UE among the UEs transmitting the same preamble. In RAPID, as $r_{ed}$ increases, the satisfiable latency requirement with each reliability slightly increases. This is because as the value of $r_{ed}$ increases, fewer PU traffic UEs transmit the same preamble on average, and thus the probability of a successful preamble transmission from at least one UE decreases.

Figs. 10(a) and 10(b) shows Empirical Cumulative Distribution Function (ECDF) of the uplink latency for smart factory applications. In Fig. 10(b), uplink latency values are more distributed between 3 ms and 4 ms than in Fig. 10(a) for all schemes. In case of RAPID, ED traffic UEs stay longer in the initial phase than PU traffic UEs. On the other hand, in the other schemes, it is possible to have the uplink packet delivered before the inactivity timer times out for ED traffic UEs. In Fig. 10(b), for ED traffic UEs, the satisfiable latency requirement with 99.999% reliability is 18.85 ms. This value is larger than the value of the PU traffic UE because fewer UEs transmit the same preamble on average, and thus the probability of a successful preamble transmission from at least one UE decreases.

Random access load: The bars in Fig. 10(c) show the sum of random access load for all applications. For each scheme, there are two types of random access load, i.e., necessary and unnecessary. The necessary random access load is the number of signals used for successful random access. In contrast, unnecessary random access load includes three types: (i) the signals when random access fails, (ii) the unnecessary RAR transmissions to the UEs who do not transmit preamble but have the same preamble ID and offset index for RAPID, and (iii) the unnecessary uplink resource allocation for RRC connection resume request messages for the UEs mentioned in (ii). In cases of 4-step RA, PRADA, and LC-MTC RA, because the ratio of PU traffic UEs who generate more packets than ED traffic UEs is reduced, the sum of random access load decreases as $r_{ed}$ increases. However, when RAPID is used for smart factory applications, the sum of random access load increases as $r_{ed}$ increases because of increasing of unnecessary RAR transmission.

The necessary random access load of PRADA is higher than that of 4-step RA, and the unnecessary random access load of PRADA is lower than that of 4-step RA. This is because PRADA allocates different random access resources to each access class, and hence, random access success ratio is higher than that of 4-step RA. LC-MTC RA further increases random access success ratio by decreasing the collision probability, thus yielding higher necessary random access load than PRADA and 4-step RA. On the other hand, unnecessary random access load of LC-MTC RA is the highest among all schemes. This is because LC-MTC RA reduces collisions at the cost of making a BS transmit separate RARs through multiple downlink control channels. In case of RAPID where $r_{ed}$ = 0.3, necessary random access load is reduced by 26.1% compared with PRADA. This is because RAPID requires only two message exchange procedures. Moreover, unnecessary random access load is decreased by 40.6% compared with PRADA. This is because UEs who perform RAPID do not suffer from random access failures due to collision. Also, with the help of APA, unnecessary RAR transmission can be minimized. Compared with PRADA, RAPID reduces the sum of random access load by 30.5% and 11.9% when $r_{ed}$ is 0.3 and 0.5, respectively. As $r_{ed}$ increases, the random access load reducing gain of RAPID decreases due to the increased unnecessary random access load, because more UEs share the same preamble and offset index. When $r_{ed}$ is 0.7, unnecessary random access load is increased further and the sum random access load of RAPID becomes comparable with that of 4step RA. RAPID, however, reduces the latency requirements that cannot be satisfied with comparison schemes.

*D. Performance of APA*

For the validation of APA performance, we consider a scenario consisting of smart factory application UEs where the number of UEs is 1,000 and $r_{ed} = 0.3$. We allocate $n(S_{cr})$ to 54, which is the maximum value defined in (32). In the initial phase, APA perfectly distinguishes the traffic type of UEs, i.e., the period of the PU traffic UEs. Fig. 11 shows the sum of random access load for smart factory applications and without APA. When APA is not applied, all UEs regardless of traffic types are uniformly allocated to the combinations of preambles and allowed slot numbers. When APA is applied, however, the UEs can be classified according to traffic types. Especially, traffic characteristics of PU traffic UEs can be grasped by APA and unnecessary random access load can be reduced by 98% compared with RAPID without APA.

VIII. CONCLUDING REMARKS

We propose RAPID, which is a new random access procedure for delay-sensitive UEs in RRC INACTIVE state introduced in 5G. RAPID completes the random access procedure by exchanging two messages using AS context ID of UE in RRC INACTIVE state. We also develop APA for reducing random access load caused by unnecessary RAR transmission. We then develop an optimization problem to obtain the number of preambles for RAPID based on random access load analysis. We also validate the analysis via comparison with simulation results. Through simulations and mathematical analysis considering mMTC devices, we demonstrate that RAPID can support delay-sensitive UEs by satisfying more strict latency requirement compared with the state-of-the-art schemes. Therefore, the proposed scheme can play important roles in satisfying the latency requirements of various applications targeted in 5G.

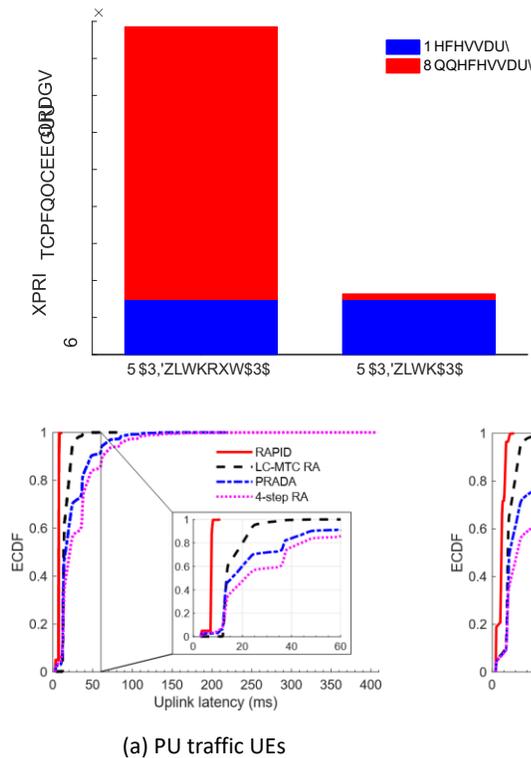

(a) PU traffic UEs  (b) ED traffic UEs  (c) Sum of random access load

Fig. 10. Two left-side graphs represent ECDF of uplink latency for smart factory applications when $N_{cb} = 23,000$, $N_{cr} = 1,000$, and $r_{ed} = 0.3$. The right-side graph shows the sum of random access load for all applications.

Fig. 11. Sum of random access load for smart factory applications with and without APA when $N_{cr} = 1,000$ and $r_{ed} = 0.3$.

PU or ED traffic. In case of PU traffic UEs, the average of estimated period is 49.99 ms, and the variance is 0.015. It can be seen that APA accurately predicts the traffic type of UEs and

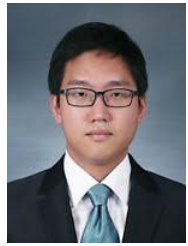
Junseok Kim (S'14) received the B.S. degree in electronics and electrical engineering from ChungAng University in 2011. He is currently pursuing the Ph.D. degree with the Department of Electrical and Computer Engineering from Seoul National University (SNU). His current research interests include 5G, Beyond 5G cellular networks, and cross-layer protocol design for wireless networks.

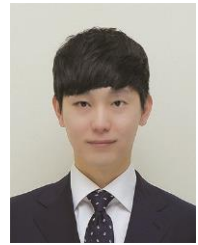
Seongwon Kim (S'12-M'18) is a Research Engineer at AI Center, SK Telecom, Seoul, Korea. Before joining SK Telecom, he was a Post-Doctoral Researcher with Seoul National University (SNU), Seoul, Korea, from Oct. 2017 to Feb. 2019. He was also a Visiting Research Scholar at the Department of Electrical and Computer Engineering, Rice University, USA, from July 2017 to Oct. 2017. He received the B.S. degree in Electrical Engineering from the Pohang University of Science and Technology (POSTECH) in 2011, and the M.S. and Ph.D. degrees in Electrical and Computer Engineering from SNU in 2013 and 2017, respectively. His current research interests include applied AI, IoT connectivity, and next-generation wireless networks.

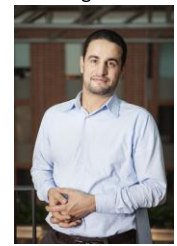
Tarik Taleb Prof. Tarik Taleb is currently Professor at the School of Electrical Engineering, Aalto University, Finland. He is the founder and director of the MOSA!C Lab (www.mosaic-lab.org). He is also working as part time professor at the Center of Wireless Communications, University of Oulu. Prior to his current academic position, he was working as Senior Researcher and 3GPP Standards Expert at NEC Europe Ltd, Heidelberg, Germany. He was then leading the NEC Europe Labs Team working on R&D projects on carrier cloud platforms. Before joining NEC and till Mar. 2009, he worked as assistant professor at the Graduate School of Information Sciences, Tohoku University, Japan. From Oct. 2005 till Mar. 2006, he worked as research fellow at the Intelligent Cosmos Research Institute, Sendai, Japan.

He received his B. E degree in Information Engineering with distinction, M.Sc. and Ph.D. degrees in Information Sciences from Tohoku Univ., in 2001, 2003, and 2005, respectively. Prof. Talebs research interests lie in the field of architectural enhancements to mobile core networks (particularly 3GPPs), network softwarization&slicing, mobile cloud networking, network function virtualization, software defined networking, mobile multimedia streaming, inter-vehicular communications, and social media networking. Prof. Taleb has been also directly engaged in the development and standardization of the Evolved Packet System as a member of 3GPP System Architecture working group. Prof. Taleb is a member of the IEEE Communications Society Standardization Program Development Board.

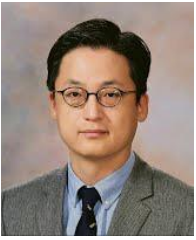

Sunghyun Choi (S'96-M'00-SM'05-F'14) is a professor at the Department of Electrical and Computer Engineering, Seoul National University (SNU), Seoul, Korea. Before joining SNU in September 2002, he was with Philips Research USA, Briarcliff Manor, New York, USA, as a Senior Member Research Staff for three years. He was also a visiting associate professor at the Electrical Engineering department, Stanford University, USA, from June 2009 to June 2010. He received his B.S. (summa cum laude) and M.S. degrees in electrical engineering from Korea Advanced Institute of Science and Technology (KAIST) in 1992 and 1994, respectively, and received Ph.D. at the Department of Electrical Engineering and Computer Science, The University of Michigan, Ann Arbor in September, 1999.

His current research interests are in the area of wireless/mobile networks with emphasis on IoT connectivity, WLAN/WPAN, next-generation mobile networks, and acoustic communication. He co-authored over 250 technical papers and a book "Broadband Wireless Access and Local Networks: Mobile WiMAX and WiFi," Artech House, 2008 (with B. G. Lee). He holds over 160 patents, and numerous patents pending. He has served as a General Co-Chair of COMSWARE 2008, a Program Committee Co-Chair of IEEE DySPAN 2018, ACM Multimedia 2007, and IEEE WoWMoM 2007. He has also served on program and organization committees of numerous leading wireless and networking conferences including ACM MobiCom, IEEE INFOCOM, IEEE SECON, and IEEE WoWMoM. He is also currently serving as an editor of IEEE Transactions on Wireless Communications, and served as an editor of IEEE Transactions on Mobile Computing, IEEE Wireless Communications Magazine, ACM SIGMOBILE Mobile Computing and Communications Review, Journal of Communications and Networks, Computer Networks, and Computer Communications. He has served as a guest editor for IEEE Journal on Selected Areas in Communications, IEEE Wireless Communications, and ACM Wireless Networks. From 2000 to 2007, he was an active contributor to IEEE 802.11 WLAN Working Group.

He has received numerous awards including KICS Dr. Irwin Jacobs Award (2013), Shinyang Scholarship Award (2011), Presidential Young Scientist Award (2008), IEEK/IEEE Joint Award for Young IT Engineer (2007), Outstanding Research Award (2008) and Best Teaching Award (2006), both from the College of Engineering, Seoul National University; the Best Paper Award from IEEE WoWMoM 2008, and Recognition of Service Award (2005, 2007) from ACM. Dr. Choi was a recipient of the Korea Foundation for Advanced Studies (KFAS) Scholarship and the Korean Government Overseas Scholarship during 1997–1999 and 1994–1997, respectively. He was named IEEE fellow in 2014 for the contribution to the development of WLAN protocols.